# OVERVIEW ON STUDIES OF MARTIAN LIKE $CO_2$-$N_2$ MIXTURE BY INDUCTIVELY COUPLED PLASMA TORCH

**Vacher D.**[(1)], **André P.**[(1)], **Dudeck M.**[(2)]

[(1)]LAEPT, CNRS, Clermont-Ferrand, France, Damien.VACHER@laept.univ-bpclermont.fr
[(2)]Institut Jean Le Rond d'Alembert, University of Paris 6, France

**ABSTRACT**

The purpose of the work is to make an overview on the results obtained through the studies on martian plasmas created by inductively coupled plasma torches (ICP). As the main advantage of the ICP torch is the absence of electrode compared to the others various test facilities, the radiative properties of this kind of plasmas are of interest to propose test cases in order to validate radiation models. ICP torches can work under various operating conditions in terms of pressure, enthalpy or flow. As a consequence, the studied plasma can be either at thermodynamical equilibrium or out of equilibrium, without problems of stability in time. The presentation concerns only the plasmas formed with a martian like $CO_2$-$N_2$ mixture and all the parameters of test facilities will be precised.

The following paper corresponds to the first step of a global paper which will be proposed later and it reports only the oral presentation which has been done during the third International Workshop of RHTG.

## 1. INTRODUCTION

Five laboratories have been found concerning the study of $CO_2$ or $CO_2$-$N_2$ plasmas by inductively coupled plasma torch. The laboratories which are about to start this kind of study do not appear in this presentation.

The laboratories which have obtained results are the following:
- L.A.E.P.T. (Laboratory of Electrical Arc and Thermal Plasmas) located in Clermont-Ferrand, France;
- C.O.R.I.A. (Complexe de Recherche Interprofessionnel en Aérothermochimie) located at Rouen, France;
- I.R.S. (Institut für RaumfahrtSysteme) located at Stuttgart, Germany;
- V.K.I. (Von Karman Institute) located at Rhode Saint Genese, Belgium;
- I.P.M. (Institute for Problems of Mechanics) located at Moscow, Russia.

## 2. L.A.E.P.T. (Clermont-Ferrand, France)

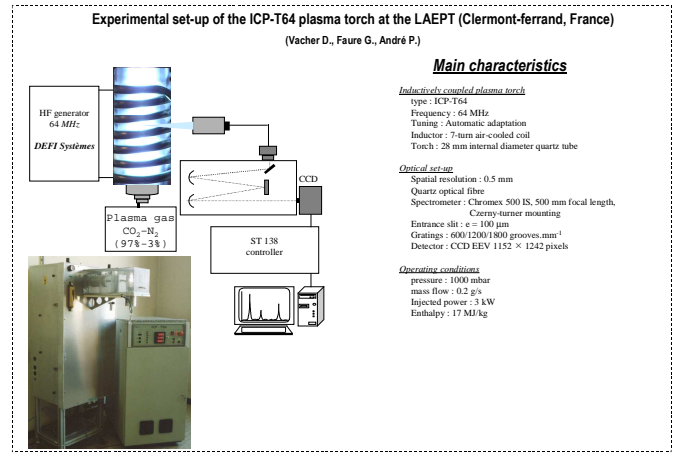

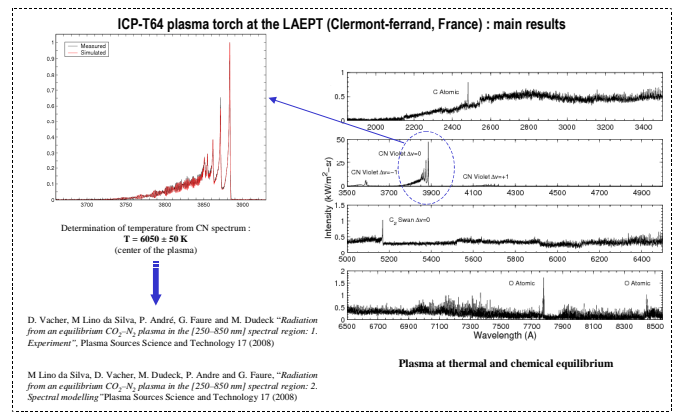

## 3. C.O.R.I.A. (Rouen, France)

### ICP torch at the CORIA (Rouen, France) : (1)
(Rond C., Bultel A., Boubert P., Chéron B.G.)

**Main characteristics**

*Inductively coupled plasma torch*
- Frequency : 13.56 MHz
- Inductor : 7-turn water-cooled coil
- Torch : 60 mm (coil location)
- 90 mm (downstream)

*Optical set-up*
- periscope system
- Spectrometer : SpectraPro-300i,
- Entrance slit : e = 50 μm
- Gratings : 1200 grooves.mm$^{-1}$ (blazed at 300 nm)
- Detector : ICCD camera PI-Max (Princeton)

*Operating conditions*
- pressure : 1-2 mbar
- flow rate : 0.1 – 0.3 slm
- Max. injected power : 2 kW
- Power delivering to the plasma : 80-160 W
- Specific enthalpy : 13.6 – 40.8 MJ/kg

This experimental set-up has been presented by A. Bultel :
«*Spatial evolution of the radiation from a non equilibrium $CO_2$ RF plasma*»

### ICP torch at the CORIA (Rouen, France) (1) : main results
(Rond C., Bultel A., Boubert P., Chéron B.G.)

*Aim of the study :*
- to analyse the $CO_2$ dissociation at low pressure
- to calculate CO+O chemiluminescent emission
- to investigate the evolution of the plasma chemistry from the creation zone to downstream
- to estimate the CO and O densities in their ground state and their excited ones

**Plasma under thermal and chemical nonequilibrium**

| | $T_r$ (K) | $T_v$ (K) | $T_{eh}$ (K) | density of the upper state (m$^{-3}$) |
|---|---|---|---|---|
| Triplet (d$^3\Delta_i \to$ a$^3\Pi_r$) | 500 | 1000 | 5000 | 10$^{15}$ |
| Asundi (a'$^3\Sigma^+ \to$ a$^3\Pi$) | 500 | 7000 | 5000 | 10$^{14}$ |

Estimation of temperatures from CO system in the post discharge

### ICP torch at the CORIA (Rouen, France) : (2)
(Boubert P.)

**Main characteristics**

*Inductively coupled plasma torch*
- Frequency : 1.76 MHz
- Inductor : 5 coils
- Torch : 100 mm diameter quartz tube

*Optical set-up*
- Spectrometer : Sopra1500,
- Entrance slit : e = 7 μm
- Gratings : 2400 grooves.mm$^{-1}$ (blazed at 400 nm)
- Detector : Photodiode array (Thomson TH7821)

*Operating conditions*
- pressure : 20 mbar
- flow rate : 18 L.min$^{-1}$
- injected power : 25 kW
- specific enthalpy : ? MJ/kg
- plasma gas : $CO_2$-$N_2$-Ar (95%-3%-2%)

| Violet system of CN | $T_{rot}$ (K) | $\Delta T_{rot}$ (K) | $T_{vib}$ (K) | $\Delta T_{vib}$ (K) |
|---|---|---|---|---|
| FWHM = 0.06 nm | 8000 | 500 | 10000 | 200 |
| FWHM = 0.08 nm | 8000 | 500 | 10000 | 200 |

| $C_2$ system | $T_{rot}$ (K) | $\Delta T_{rot}$ (K) | $T_{vib}$ (K) | $\Delta T_{vib}$ (K) |
|---|---|---|---|---|
| FWHM = 0.075 nm | 9000 | 1000 | 10000 | 1000 |

Estimation of temperatures

Conclusion of the study :
- CN : good accuracy experiment-theory
- $C_2$ : notables differences between experiment and theory
- strong perturbations of states
- anomalies of states population

## 4. I.R.S. (Stuttgartt, Germany)

### ICP torch at the IRS (Stuttgart, Germany)
(Herdrich G., Endlich P. and al.)

**Main characteristics**

*Inductively heated generator IPG4*
- Frequency : 0.6 MHz
- Inductor : 5-turn water-cooled coil
- Torch : 84mm internal diameter quartz tube
- Nozzle throat diameter : 50 mm

*Optical set-up*
- mini-Spectrometer :
- Entrance slit : e = ? μm
- Gratings : ?
- Detector : 2000 pixel detector (pixel resolution : 0.3 nm)

*Operating conditions*
- pressure : 1.9-8 mbar
- $CO_2$ mass flow : 3.7 g.s$^{-1}$
- $N_2$ mass flow : 70 g.s$^{-1}$
- Thermal plasma power : 20 kW

*Measurements in the plasma :*
- Heat flux
- Pitot pressure
- Thermal plasma power (calorimeter)
- **Study of the influence of dust particles**
(injection of iron oxides and silicon dioxides with φ < 10μm)

View of plasma source IPG4

P = 500 Pa

### ICP torch at the IRS (Stuttgart, Germany) : Example of spectra
(Herdrich G., Endlich P. and al.)

P = 190 Pa — Spectra of boundary layer of the sample / Spectra of free stream
P = 800 Pa — Spectra of boundary layer of the sample / Spectra of free stream

*No values of the estimated temperatures of plasma*

### ICP torch at the IRS (Stuttgart, Germany) : First estimation of temperatures
(Lein S., Herdrich G. and al.)

The results are issued from a poster presented in the 6$^{th}$ International Planetary Probe workshop (2008)
« *Characterization of $CO_2$ plasma free stream conditions for atmospheric entry simulation* »

- pure $CO_2$ plasma
- ambient pressure (1.3 mbar)
- mass flow : 2.2 g.s$^{-1}$
- anode power : 125 kW

Fabry-Perot interferometry

Optical emission spectroscopy – PARADE radiation simulation

First Conclusion : disagreement theory–experimental data (quantitatively not qualitatively)

## 5. V.K.I. (Rhode Saint Genese, Belgium)

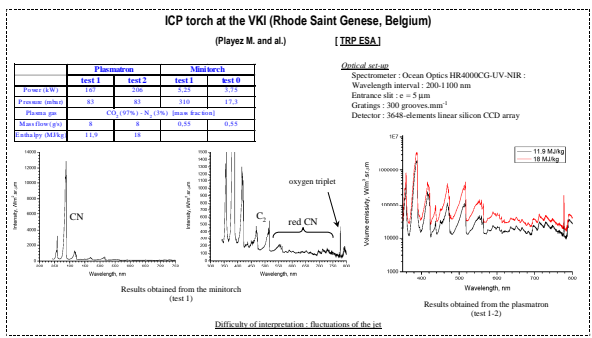

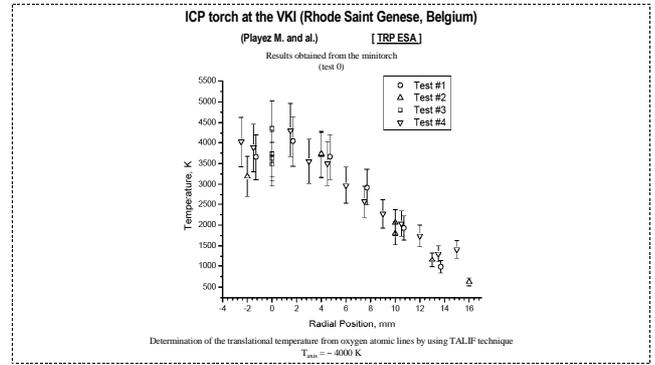

## REFERENCES

The informations are given by M. Playez.

## 6. I.P.M. (Moscow, Russia)

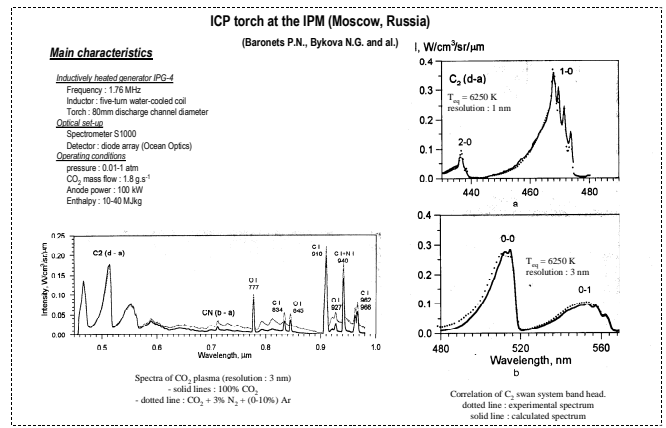

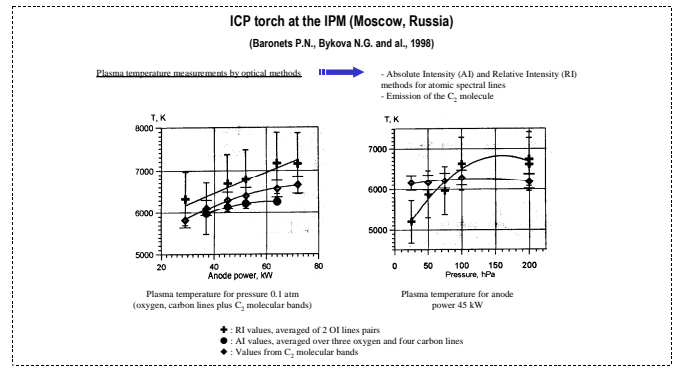

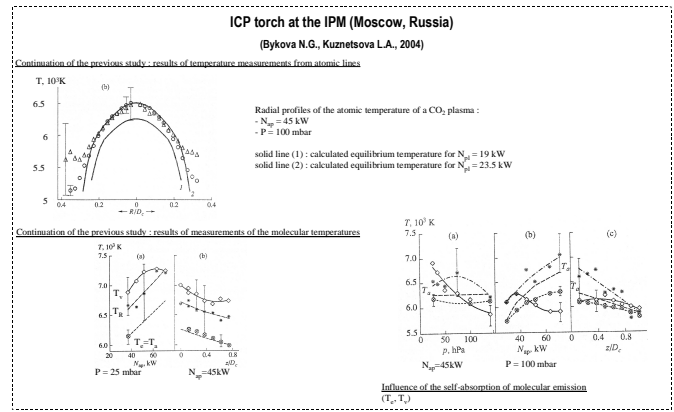

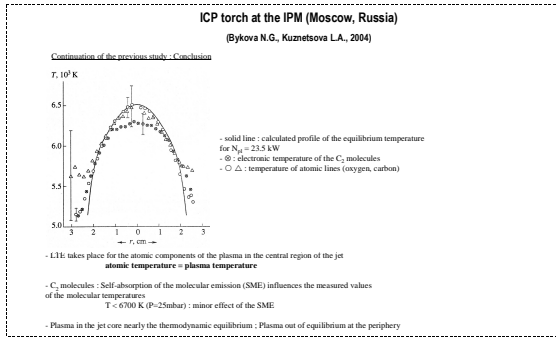

## 7. CONCLUSION

A conclusion can not be done due to the fact that each laboratory works under different conditions in terms of flow, injected power, operating frequency and operating pressure. However, it is very interesting to gather all the data in order the domain of applications of each laboratory and also in order to create a connection between all these laboratories.

The following figure propose a regrouping of the data of each laboratory in terms of specific enthalpy and pressure. Of course, The field of investigation of each laboratories is likely to evolve.

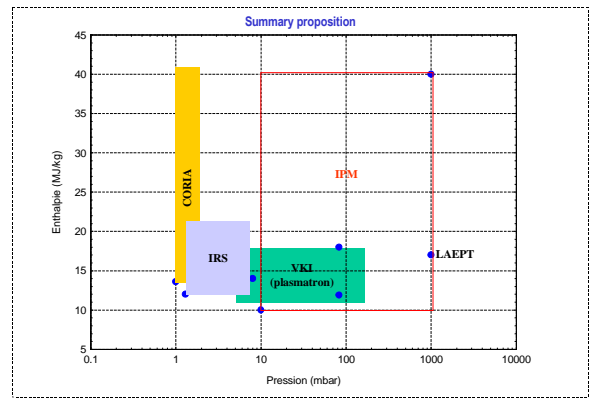

## ACKNOWLEDGMENTS

I would like to thank all the colleagues who permitted to this presentation to exist. I express all my recognition to them and thanks once again them for all the informations that they sent to me. These colleagues are : P. Boubert, A. Bultel, G. Herdrich, A.F. Kolesnikov, M. Playez and C. Rond.

I wish that the discussion will continue in order to still work together on the atmospheric entries with inductively coupled plasma torches.